\documentclass[journal=jacsat,manuscript=article]{achemso}

\usepackage{chemformula} 
\usepackage[T1]{fontenc} 
\usepackage{siunitx}
\usepackage{gensymb}

\author{M. D'Amato}
\affiliation[LKB]{Laboratoire Kastler Brossel, Sorbonne Universit\'e, CNRS, ENS-PSL Research University, Coll\`ege de France, 4 place Jussieu, 75252 Paris Cedex 05, France}
\author{Ningyuan Fu}
\affiliation[ESPSI]{Laboratoire de Physique et d’Etude des Matériaux, ESPCI-Paris, PSL Research University, Sorbonne Université Univ Paris 06, CNRS UMR 8213, 10 rue Vauquelin 75005 Paris, France}
\affiliation[INSP]{Sorbonne Université, CNRS , Institut des NanoSciences de Paris, 4 place Jussieu, 75005 Paris, France}
\author{Quentin Glorieux}
\affiliation[LKB]{Laboratoire Kastler Brossel, Sorbonne Universit\'e, CNRS, ENS-PSL Research University, Coll\`ege de France, 4 place Jussieu, 75252 Paris Cedex 05, France}
\author{Elisabeth Giacobino}
\affiliation[LKB]{Laboratoire Kastler Brossel, Sorbonne Universit\'e, CNRS, ENS-PSL Research University, Coll\`ege de France, 4 place Jussieu, 75252 Paris Cedex 05, France}
\author{Hanna Le Jeannic}
\affiliation[LKB]{Laboratoire Kastler Brossel, Sorbonne Universit\'e, CNRS, ENS-PSL Research University, Coll\`ege de France, 4 place Jussieu, 75252 Paris Cedex 05, France}
\author{Sandrine Ithurria}
\affiliation[ESPSI]{Laboratoire de Physique et d’Etude des Matériaux, ESPCI-Paris, PSL Research University, Sorbonne Université Univ Paris 06, CNRS UMR 8213, 10 rue Vauquelin 75005 Paris, France}
\author{Emmanuel Lhuillier}
\affiliation[INSP]{Sorbonne Université, CNRS , Institut des NanoSciences de Paris, 4 place Jussieu, 75005 Paris, France}
\author{Alberto Bramati}
\affiliation{Laboratoire Kastler Brossel, Sorbonne Universit\'e, CNRS, ENS-PSL Research University, Coll\`ege de France, 4 place Jussieu, 75252 Paris Cedex 05, France}
\email{alberto.bramati@lkb.upmc.fr}

\title[An \textsf{achemso} demo]
  {Room-temperature efficient single-photon generation from CdSe/ZnS nanoplateletes}

\abbreviations{IR,NMR,UV}

\begin{document}


\begin{abstract}
In the search for materials for quantum information science applications, colloidal semiconductor nanoplatelets (NPLs) have emerged as a highly promising new class of materials due to their interesting optical properties, such as narrow emission linewidth and fast photoluminescence (PL) lifetimes at room temperature. So far only few works focused on the quantum properties of their emission, however, NPLs, with their atomic-scale thickness and one-dimensional quantum confinement, are promising candidates for single-photon sources. Here, we demonstrate room-temperature single-photon emission from core/shell CdSe/ZnS NPLs, which feature 8 × 20 \ch{nm^2} surface area and 1 nm shell. The limited surface area ensures effective Auger non-radiative recombination, resulting in highly efficient single-photon generation with values of photon purity as low as $g^{(2)}(\tau)=0.04$. The observed long-period blinking and bleaching, typical of such thin shells, can be easily reduced by increasing the shell thickness. This work establishes NPLs as new single-photon sources very well suited for integration into quantum photonic systems.
\end{abstract}
\section{Introduction}

The quest for an optimal single-photon source (SPS) with superior emission properties lies at the heart of optical quantum-information science research. Over the years, various quantum systems have been explored for their ability to deliver on-demand optical coherent single photons, including atoms and ions\cite{kuhn2002deterministic}, organic molecules\cite{toninelli2021single}, defects in diamonds\cite{schroder2011ultrabright}, and semiconductor atom-like systems\cite{aharonovich2016solid}. Among these, semiconductor quantum emitters, especially III-V epitaxially self-assembled quantum dots (QDs), have emerged as a prominent solid-state platform for developing quantum devices\cite{senellart2017high}.

Concurrently, colloidal nanocrystals (NCs) have recently gained significant attention due to their ability to offer high-quality single-photon emission at room temperature and easy integration into quantum photonics devices. Colloidal NCs are synthesized in a wide variety of compositions, sizes and shapes. Indeed, their family encompasses QDs (with 3D confinement), nanowires (with 2D confinement), and nanoplatelets (NPLs) (with 1D confinement). NPLs feature a ultra-thin thickness precisely adjusted by controlling the number of atomic monolayers, as well as a controllable lateral size. These unique characteristics endow NPLs with remarkable optical properties, including a large exciton binding energy, significant oscillator strength, narrow emission linewidths, fast photoluminescence (PL) lifetimes, large absorption cross-section, and high optical gain\cite{yu2020optical}.

Exploiting these interesting attributes, NPL ensembles have found several applications in the optoelectronic domain such as light-emitting diodes (LEDs), lasing, and luminescent solar concentrators (LSCs)\cite{diroll20232d}. Our focus here is on advancing their quantum functionality at the single-particle level by demonstrating efficient room-temperature single-photon generation from core/shell CdSe/ZnS NPLs.

\section{Results and discussion} 

\begin{figure*}[ht!]
     \includegraphics[width=140 mm]{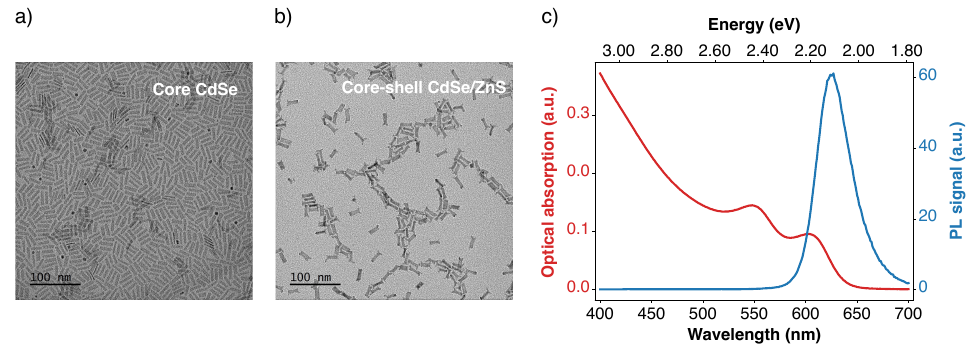}
     \caption{ CdSe/ZnS NPLs.
      a) Transmission electron microscopy image of the core CdSe NPLs. b) Transmission electron microscopy image of the CdSe/ZnS NPLs after shell growing. c) Absorption and photoluminescence spectra of the CdSe/ZnS NPLs.}
   \label{fig:FIG1}
\end{figure*}

The initial core CdSe NPLs were synthesized in order to minimize their planar extension to approximately 8 × 20 \ch{nm^2} (see Section S1 in Supporting Informations for details), as measured from TEM image in Figure \ref{fig:FIG1}a. The size minimization maintains the colloidal stability of the NPLs and it is critical to make them behave as single photon emitters. 
Unlike NPLs with larger lateral dimensions that tend to aggregate in solution, this choice prevents NPLs from stacking\cite{abecassis2014self}, which is difficult to break down with sonification. Instead, these NPLs with smaller surface can be dispersed in hexane without noticeable aggregation. This property is later critical to adress individual particles. The synthesized core NPLs have a thickness of 1.2 nm.
Due to the resulting high aspect-ratio, the width acts as the confinement-inducing dimension. We then overcoat the NPLs with a thin 1 nm ZnS shell (Figure  \ref{fig:FIG1}b) to obtain core/shell CdSe/ZnS NPLs with type-I band alignment, which exhibit both narrow linewidths and high PLQYs\cite{zhang2023visible,kelestemur2019colloidal}. The PL and absorbance spectra of an ensemble of CdSe/ZnS core/shell NPLs are illustrated in Figure \ref{fig:FIG1}c. The ZnS shelling shifts the band edge emission from 510 nm\cite{zhang2023visible} to approximately 630 nm. The peak emission wavelength and the full-width-half-maximum (FWHM) of the PL emission are 627 nm (1.98 eV) and 39 nm (123 meV), respectively. The quantum efficiency of these NPLs, around $35\%$ \cite{qu2020nanoplatelet}, and their good dispersion stability enabled single-particle spectroscopy, which was previously challenging for probing the single-photon character of their emission.

\begin{figure*}[ht!]
     \includegraphics[width=120 mm]{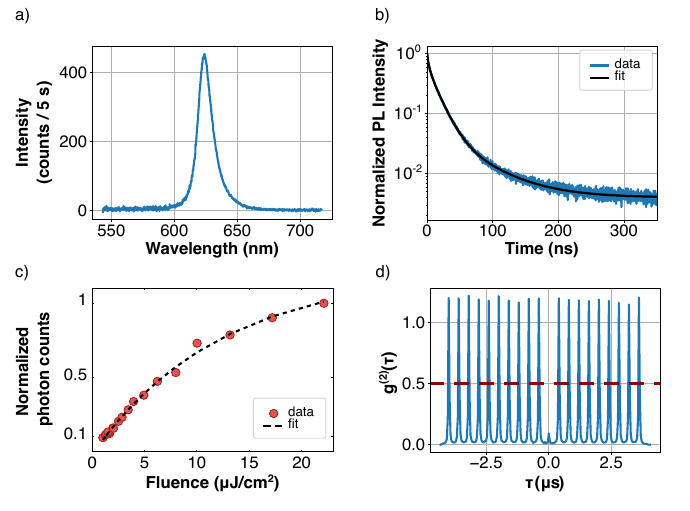}
     \caption{ Single photon emission from an individual CdSe/ZnS NPL.
      a) Typical PL emission spectrum of a CdSe/ZnS NPL. b)  PL decay of a CdSe/ZnS NPL. c) Saturation measurement of a individual CdSe/ZnS NPL. The red dots are the measured counts with the excitation power. The dashed black line is the fitting function from eq. \ref{eq:saturation}. d) Second-order correlation function of a single CdSe/ZnS NPL emitting high quality single photons ( $g^{(2)}(0)=0.04$ after background counts subtraction). The red dashed line indicates the threshold for the single-photon emission claim.}
   \label{fig:FIG2}
\end{figure*}

To study this, the NPLs were diluted in hexane, drop-casted on a microscope cover slide and their fluorescence observed with an inverted confocal microscope (see Methods for details on the experimental setup). The typical PL emission spectrum of a single NPL at room temperature is presented in Figure \ref{fig:FIG2}a. In comparison to ensemble measurement, the spectral width is reduced to 14 nm (42.52 meV) and the central emission wavelength (CEW) is shifted to 624 nm (1.99 eV) at room temperature. The full-width at half maximum (FWHM) of these NPLs is even smaller to the one of QDs of the same material \cite{zhang2023visible, huang2021synthesis} or other high-quality colloidal nanocrystals (NCs) used for single photon generation at room temperature \cite{raino2022ultra}. 

The emission decay is plotted in Figure \ref{fig:FIG2}b and fitted with a tri-exponential model, giving lifetimes (amplitudes) of 2.8 ns ( $30\%$), 15.3 ns ($65\%$) and 56 ns ($5\%$). We attribute the last long component to the background noise. 
Figure \ref{fig:FIG2}c shows the PL intensity as a function of the excitation fluence. The PL intensity of the CdSe/ZnS NPLs first increases with the excitation power. Then, as expected for colloidal NCs which are not perfect two-level systems, it saturates according to the relation:
\begin{equation}
\label{eq:saturation}
    I=A\cdot \bigg[ 1- e^{-\frac{P}{P_{sat}}}\bigg] + B \cdot  P
\end{equation}
where \ch{P_{sat}} is the saturation power (i.e. the excitation power at which the number of excitons \ch{N_{eh}= 1}). A and B are two constants that take account of the single- and bi-exciton components of the emission, respectively.
Fitting the experimental data with this model, the extracted saturation power $P_{sat}$ corresponds to a fluence of around $9$ $\mu J/cm^2 $. The time-resolved measurements here reported are taken below saturation, specifically at $0.25 \cdot P_{sat}$. Even at low excitation power, the high brightness of these NPLs guarantees an optimal signal/noise ratio. 

We then investigated if CdSe/ZnS NPLs can serve as a source of quantum light, henceforth measuring the second-order correlation function $g^{(2)}(\tau)$ in a Hanbury-Brown and Twiss set-up with pulsed excitation and in time-tagged time-resolved (TTTR) mode (see Methods for details).
We obtain values of photon purity as low as $g^{(2)}(\tau)=0.04$ (after background subtraction), 
 unequivocally proving that CdSe/ZnS NPLs can serve as high quality single photon sources.
 
Moreover, this is so far, to the best of our knowledge, the lowest reported values for CdSe-based NPLs\cite{liu2020fourier,benjamin2020temperature,ma2017size} as well as for core-shell QDs\cite{proppe2023highly,jiang2021single} at room temperature. In the CdSe/ZnS core/shell QDs single photon emission is attributed to highly efficient, non-radiative Auger recombination of multi-excitons\cite{michler2000quantum,lounis2000photon}. In transitioning from 0D to 2D systems, the weak in-plane confinement facilitates the spatial spreading of excitons in the NPL lateral plane, potentially reducing Auger recombination effects. However, Auger recombination time has been shown to depend on lateral area and thickness\cite{li2017area} and the small lateral area of our NPLs may ensure efficient suppression of multi-excitons.

\begin{figure*}[ht!]
     \includegraphics[width=140 mm]{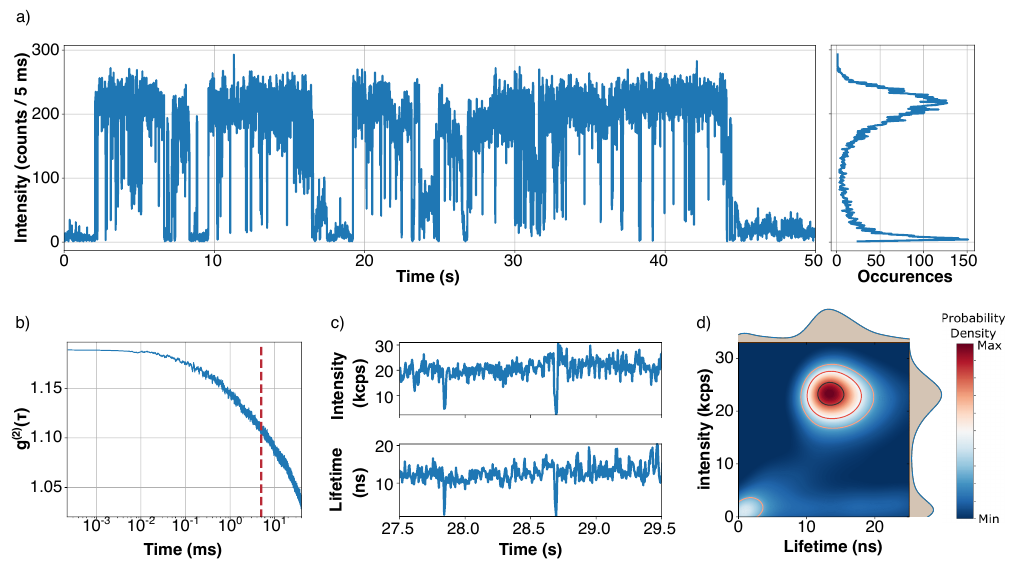}
     \caption{ Blinking and bleaching in a single CdSe/ZnS NPL. 
      a) PL time-trace of a single NPL with a bin time of 5 ms and corresponding histogram of intensity occurrences. b) $g^{(2)}(\tau)$ for time delays between $1 \mu s$ and 1s. The red-line at 5 ms corresponds to the choosen bin time. c) Zoom-view of the PL trajectory (upper box) and lifetime trajectory (lower box). d)Fluorescence lifetime-intensity distribution (FLID) image with a 5 ms bin time. }
   \label{fig:FIG3}
\end{figure*}

 We now turn to further characterize the emission properties of our NPLs. The PL time-trace of a single NPL, below saturation, is displayed in Figure \ref{fig:FIG3}a for the first 50 s (see Section S3 in Supporting Information for the entire trace) and with a binning time of 5 ms. Emission intermittency, with switching between high and low-intensity states, can be observed and is confirmed in the histogram of emission (Figure \ref{fig:FIG3}a) where two intensity peaks can clearly be identified. This phenomenon called blinking is common for nanoscale colloidal emitters\cite{moerner1997those,bradac2010observation,efros2016origin,frantsuzov2008universal}, and can typically hamper their use as single photon emitters. 
 
To evaluate the typical timescale of the blinking, we measured the second order correlation function $g^{(2)}(\tau)$ and plotted the envelope of its coincidence peaks for delays between approximately 1 us and 100 ms (see Figure \ref{fig:FIG3}b). We can interpret the $g^{(2)}(\tau)$ as quantifying how similar the emitter's state at time $t_0 + \tau $ is to its state at time $t_0$. If the delay $\tau $ falls within the timescale of intensity intermittency, a bunching ($g^{(2)}(\tau) > 1$) is visible\cite{rabouw2019microsecond,messin2001bunching}. 
Compared to methods relying on binning the photon detection events, this method does not suffer from any a priori assumptions on the blinking rate, and yields information on timescales as short as $\mu s$ \cite{d2023highly}. 
Here, the $g^{(2)}(\tau)$ function envelope has a bunching value of 1.18 at 1 $\mu s$ and then slowly decreases towards unity for delays higher than approximately 100 $\mu s$, indicating blinking over a wide timescale compared to other colloidal nanocrystals with fast blinking dynamics\cite{manceau2018cdse,d2023highly}. Such long blinking characteristic time might be decreased by increasing the shell thickness \cite{chen2008giant,mahler2008towards}, currently limited by the mismatch between CdSe and ZnS that challenge the growth of a thick shell.

To delve into the blinking mechanism of CdSe/ZnS NPLs, we can analyze the link between their PL intensity and photon lifetime. Auger blinking involves colloidal emitters transitioning into a charged state (trion) during constant photo-excitation, leading to rapid nonradiative recombination where excess charge carriers transfer exciton energy without emitting photons. This diminishes PL intensity and photon lifetime until neutralization of excess charge. Surface-trap-induced blinking, differently, occurs due to non-radiative relaxation via surface traps, causing random blinking without intensity lifetime correlation.
 
In Figure \ref{fig:FIG3}c, we show a zoom in PL intensity trace and the corresponding lifetimes, with 5ms binning time. The two traces are clearly correlated. 
To further assess the origin of our NPL blinking process, we employed a fluorescence lifetime-intensity distribution (FLID) analysis over the 50 s trace of Figure \ref{fig:FIG3}a, which is an effective experimental approach to discriminate the origin of the blinking \cite{galland2012lifetime,trinh2020verification,rabouw2019microsecond,pierini2020highly}. 
The overall density of occurrence bins with a given pair of intensity (y-axis) and lifetime (x-axis), is displayed in Figure \ref{fig:FIG3}d.
Such FLID image clearly confirms that two different states, connected with a curved line, contribute to the PL, a signature of Auger blinking \cite{kim2019elucidation,trinh2020verification}. 
   Note that the pattern near low intensity in Figure \ref{fig:FIG3}d is caused by the background noise \cite{yuan2018two}. 

\section{Methods}

\textbf{CdSe/ZnS core-shell NPLs.} 
We prepared 4.5 monolayer (ML) thick CdSe nanoplatelets (NPLs), consisting in 4 Se crystalline planes bound to 5 planes of Cd following the procedure outlined in Zhang et al.\cite{zhang2023visible}. See details about the synthesis in Supporting Information.\newline
 \textbf{Optical and Structural Characterization of Colloidal NPLs Ensembles.} \newline
\textit{Absorption and PL spectra.} 
UV-visible spectra are acquired with a Cary 5000 spectrometer. Photoluminescence and excitation spectra are obtained with an Edinburgh Instrument spectrometer. During the measurements, the NPLs are dispersed in hexane. For PL excitation measurements, care has been taken to work in diluted solution to limit emission reabsorption phenomena.
\textit{Transmission Electron Microscopy.} A drop of diluted QDs solution was drop-casted onto a copper grid covered with an amorphous carbon film. The grid was degassed overnight under secondary vacuum. Imaging was conducted using a JEOL 2010 transmission electron microscope operated at 200 kV.  \newline
\textbf{Optical Characterization of NPLs at single-particle level} \newline
\textit{Steady-state microscopy.} 
The optical measurements were performed using a microscope equipped with a motorized stage with high-resolution x-y plane displacement, enabling the investigation of the entire sample. Firstly, a light-emitting diode (LED) light beam at 400 nm (CooLED pE-100) in wide-field configuration is employed to illuminate a large area of the sample and to facilitate the systematic selection of individual emitters.
Then, a pulsed laser (LD405-B, PicoQuant) in confocal configuration is employed to excite an individual targeted NPL. In both schemes, the excitation light is reflected towards the sample using a dichroic mirror placed at an incident angle of \SI{45}{\degree} to the incoming light.  A high numerical aperture (NA $=$ 1.4) infinity-corrected oil objective with 100X magnification is used to both focus the excitation light and back-collect the NPL’s fluorescence, which is filtered to remove spurious laser excitation. For steady-state measurements, the collected light can be imaged on a camera (HAMAMATSU ORCA Flash 4.0 LT) or sent to a spectrometer (Princeton Instrument Acton SP 2500) equipped with a CCD camera (Princeton Instrument Pixis 400). \newline
\textit{Time-correlated single-photon counting (TCSPC) measurements.} 
The collected PL is sent to a pair of avalanche photodiodes (APDs) (Excelitas SPCM- AQRH-14-FC) in a Hanbury Brown-Twiss (HBT) configuration. The digital output signal is sent to a router (PDL 800, Picoquant) which direct them to a TCSPC card (PicoHarp 300, Picoquant) working in time-tagged time-resolved (TTTR) mode.  

 \section{Conclusion}

 In conclusion, we conducted an optical spectroscopy study on core/shell CdSe/ZnS NPLs at the single-particle level, demonstrating for the first time their single-photon emission. Due to their high brightness and superior quality of the single-photon emission, these materials hold great potential for integration into photonic devices. To improve their photostability, future studies could focus on optimizing their blinking and bleaching behavior with thicker shells and investigating how Auger recombination efficiency is affected\cite{ma2017size}. Additionally, the assessment of their photoluminescence features at cryogenic temperature, including the emission linewidth, fast decay rates, spectral diffusion, and degree of polarization \cite{li2024polarized}, could further highlight their potential for optical and quantum technology.

\begin{acknowledgement}
This work received support from the French National Research Agency through the grant IPER-Nano2 (ANR-18CE30-0023) and Bright (ANR-21-CE24-0012-02), from the France 2030 program through the project OQuLus (ANR-23-PETQ-0013), by the European Union’s Horizon 2020 research and innovation program under grant agreement No 828972-Nanobright and by the ERC grant Ne2dem (grant 853049). AB and QG are members of the Institut Universitaire de France (IUF).

The authors declare no competing financial interest.

\end{acknowledgement}

\begin{suppinfo}

The following files are available free of charge.
\begin{itemize}
  \item Supporting Information: Synthesis of the CdSe/ZnS NPLs, Wide-field and confocal microscopy at single-particle level, PL time-trace.
\end{itemize}

\end{suppinfo}

\bibliography{Manuscript}

\providecommand{\latin}[1]{#1}
\makeatletter
\providecommand{\doi}
  {\begingroup\let\do\@makeother\dospecials
  \catcode`\{=1 \catcode`\}=2 \doi@aux}
\providecommand{\doi@aux}[1]{\endgroup\texttt{#1}}
\makeatother
\providecommand*\mcitethebibliography{\thebibliography}
\csname @ifundefined\endcsname{endmcitethebibliography}  {\let\endmcitethebibliography\endthebibliography}{}
\begin{mcitethebibliography}{38}
\providecommand*\natexlab[1]{#1}
\providecommand*\mciteSetBstSublistMode[1]{}
\providecommand*\mciteSetBstMaxWidthForm[2]{}
\providecommand*\mciteBstWouldAddEndPuncttrue
  {\def\EndOfBibitem{\unskip.}}
\providecommand*\mciteBstWouldAddEndPunctfalse
  {\let\EndOfBibitem\relax}
\providecommand*\mciteSetBstMidEndSepPunct[3]{}
\providecommand*\mciteSetBstSublistLabelBeginEnd[3]{}
\providecommand*\EndOfBibitem{}
\mciteSetBstSublistMode{f}
\mciteSetBstMaxWidthForm{subitem}{(\alph{mcitesubitemcount})}
\mciteSetBstSublistLabelBeginEnd
  {\mcitemaxwidthsubitemform\space}
  {\relax}
  {\relax}

\bibitem[Kuhn \latin{et~al.}(2002)Kuhn, Hennrich, and Rempe]{kuhn2002deterministic}
Kuhn,~A.; Hennrich,~M.; Rempe,~G. Deterministic single-photon source for distributed quantum networking. \emph{Physical review letters} \textbf{2002}, \emph{89}, 067901\relax
\mciteBstWouldAddEndPuncttrue
\mciteSetBstMidEndSepPunct{\mcitedefaultmidpunct}
{\mcitedefaultendpunct}{\mcitedefaultseppunct}\relax
\EndOfBibitem
\bibitem[Toninelli \latin{et~al.}(2021)Toninelli, Gerhardt, Clark, Reserbat-Plantey, G{\"o}tzinger, Ristanovi{\'c}, Colautti, Lombardi, Major, Deperasi{\'n}ska, \latin{et~al.} others]{toninelli2021single}
Toninelli,~C.; Gerhardt,~I.; Clark,~A.; Reserbat-Plantey,~A.; G{\"o}tzinger,~S.; Ristanovi{\'c},~Z.; Colautti,~M.; Lombardi,~P.; Major,~K.; Deperasi{\'n}ska,~I.; others Single organic molecules for photonic quantum technologies. \emph{Nature Materials} \textbf{2021}, \emph{20}, 1615--1628\relax
\mciteBstWouldAddEndPuncttrue
\mciteSetBstMidEndSepPunct{\mcitedefaultmidpunct}
{\mcitedefaultendpunct}{\mcitedefaultseppunct}\relax
\EndOfBibitem
\bibitem[Schr{\"o}der \latin{et~al.}(2011)Schr{\"o}der, G{\"a}deke, Banholzer, and Benson]{schroder2011ultrabright}
Schr{\"o}der,~T.; G{\"a}deke,~F.; Banholzer,~M.~J.; Benson,~O. Ultrabright and efficient single-photon generation based on nitrogen-vacancy centres in nanodiamonds on a solid immersion lens. \emph{New Journal of Physics} \textbf{2011}, \emph{13}, 055017\relax
\mciteBstWouldAddEndPuncttrue
\mciteSetBstMidEndSepPunct{\mcitedefaultmidpunct}
{\mcitedefaultendpunct}{\mcitedefaultseppunct}\relax
\EndOfBibitem
\bibitem[Aharonovich \latin{et~al.}(2016)Aharonovich, Englund, and Toth]{aharonovich2016solid}
Aharonovich,~I.; Englund,~D.; Toth,~M. Solid-state single-photon emitters. \emph{Nature photonics} \textbf{2016}, \emph{10}, 631--641\relax
\mciteBstWouldAddEndPuncttrue
\mciteSetBstMidEndSepPunct{\mcitedefaultmidpunct}
{\mcitedefaultendpunct}{\mcitedefaultseppunct}\relax
\EndOfBibitem
\bibitem[Senellart \latin{et~al.}(2017)Senellart, Solomon, and White]{senellart2017high}
Senellart,~P.; Solomon,~G.; White,~A. High-performance semiconductor quantum-dot single-photon sources. \emph{Nature nanotechnology} \textbf{2017}, \emph{12}, 1026--1039\relax
\mciteBstWouldAddEndPuncttrue
\mciteSetBstMidEndSepPunct{\mcitedefaultmidpunct}
{\mcitedefaultendpunct}{\mcitedefaultseppunct}\relax
\EndOfBibitem
\bibitem[Yu and Chen(2020)Yu, and Chen]{yu2020optical}
Yu,~J.; Chen,~R. Optical properties and applications of two-dimensional CdSe nanoplatelets. \emph{InfoMat} \textbf{2020}, \emph{2}, 905--927\relax
\mciteBstWouldAddEndPuncttrue
\mciteSetBstMidEndSepPunct{\mcitedefaultmidpunct}
{\mcitedefaultendpunct}{\mcitedefaultseppunct}\relax
\EndOfBibitem
\bibitem[Diroll \latin{et~al.}(2023)Diroll, Guzelturk, Po, Dabard, Fu, Makke, Lhuillier, and Ithurria]{diroll20232d}
Diroll,~B.~T.; Guzelturk,~B.; Po,~H.; Dabard,~C.; Fu,~N.; Makke,~L.; Lhuillier,~E.; Ithurria,~S. 2D II--VI semiconductor nanoplatelets: From material synthesis to optoelectronic integration. \emph{Chemical Reviews} \textbf{2023}, \emph{123}, 3543--3624\relax
\mciteBstWouldAddEndPuncttrue
\mciteSetBstMidEndSepPunct{\mcitedefaultmidpunct}
{\mcitedefaultendpunct}{\mcitedefaultseppunct}\relax
\EndOfBibitem
\bibitem[Ab{\'e}cassis \latin{et~al.}(2014)Ab{\'e}cassis, Tessier, Davidson, and Dubertret]{abecassis2014self}
Ab{\'e}cassis,~B.; Tessier,~M.~D.; Davidson,~P.; Dubertret,~B. Self-assembly of CdSe nanoplatelets into giant micrometer-scale needles emitting polarized light. \emph{Nano letters} \textbf{2014}, \emph{14}, 710--715\relax
\mciteBstWouldAddEndPuncttrue
\mciteSetBstMidEndSepPunct{\mcitedefaultmidpunct}
{\mcitedefaultendpunct}{\mcitedefaultseppunct}\relax
\EndOfBibitem
\bibitem[Zhang \latin{et~al.}(2023)Zhang, Guilloux, Bossavit, Fu, Dabard, Cavallo, Dang, Khalili, Abadie, Alchaar, \latin{et~al.} others]{zhang2023visible}
Zhang,~H.; Guilloux,~V.; Bossavit,~E.; Fu,~N.; Dabard,~C.; Cavallo,~M.; Dang,~T.~H.; Khalili,~A.; Abadie,~C.; Alchaar,~R.; others Visible and Infrared Nanocrystal-based Light Modulator with CMOS Compatible Bias Operation. \emph{ACS photonics} \textbf{2023}, \emph{10}, 430--436\relax
\mciteBstWouldAddEndPuncttrue
\mciteSetBstMidEndSepPunct{\mcitedefaultmidpunct}
{\mcitedefaultendpunct}{\mcitedefaultseppunct}\relax
\EndOfBibitem
\bibitem[Kelestemur \latin{et~al.}(2019)Kelestemur, Shynkarenko, Anni, Yakunin, De~Giorgi, and Kovalenko]{kelestemur2019colloidal}
Kelestemur,~Y.; Shynkarenko,~Y.; Anni,~M.; Yakunin,~S.; De~Giorgi,~M.~L.; Kovalenko,~M.~V. Colloidal CdSe quantum wells with graded shell composition for low-threshold amplified spontaneous emission and highly efficient electroluminescence. \emph{ACS nano} \textbf{2019}, \emph{13}, 13899--13909\relax
\mciteBstWouldAddEndPuncttrue
\mciteSetBstMidEndSepPunct{\mcitedefaultmidpunct}
{\mcitedefaultendpunct}{\mcitedefaultseppunct}\relax
\EndOfBibitem
\bibitem[Qu \latin{et~al.}(2020)Qu, Rastogi, Gr{\'e}boval, Livache, Dufour, Chu, Chee, Ramade, Xu, Ithurria, \latin{et~al.} others]{qu2020nanoplatelet}
Qu,~J.; Rastogi,~P.; Gr{\'e}boval,~C.; Livache,~C.; Dufour,~M.; Chu,~A.; Chee,~S.-S.; Ramade,~J.; Xu,~X.~Z.; Ithurria,~S.; others Nanoplatelet-based light-emitting diode and its use in all-nanocrystal LiFi-like communication. \emph{ACS applied materials \& interfaces} \textbf{2020}, \emph{12}, 22058--22065\relax
\mciteBstWouldAddEndPuncttrue
\mciteSetBstMidEndSepPunct{\mcitedefaultmidpunct}
{\mcitedefaultendpunct}{\mcitedefaultseppunct}\relax
\EndOfBibitem
\bibitem[Huang \latin{et~al.}(2021)Huang, Ye, Yang, Li, Qin, and Peng]{huang2021synthesis}
Huang,~L.; Ye,~Z.; Yang,~L.; Li,~J.; Qin,~H.; Peng,~X. Synthesis of colloidal quantum dots with an ultranarrow photoluminescence peak. \emph{Chemistry of Materials} \textbf{2021}, \emph{33}, 1799--1810\relax
\mciteBstWouldAddEndPuncttrue
\mciteSetBstMidEndSepPunct{\mcitedefaultmidpunct}
{\mcitedefaultendpunct}{\mcitedefaultseppunct}\relax
\EndOfBibitem
\bibitem[Rain{\`o} \latin{et~al.}(2022)Rain{\`o}, Yazdani, Boehme, Kober-Czerny, Zhu, Krieg, Rossell, Erni, Wood, Infante, \latin{et~al.} others]{raino2022ultra}
Rain{\`o},~G.; Yazdani,~N.; Boehme,~S.~C.; Kober-Czerny,~M.; Zhu,~C.; Krieg,~F.; Rossell,~M.~D.; Erni,~R.; Wood,~V.; Infante,~I.; others Ultra-narrow room-temperature emission from single CsPbBr3 perovskite quantum dots. \emph{Nature communications} \textbf{2022}, \emph{13}, 2587\relax
\mciteBstWouldAddEndPuncttrue
\mciteSetBstMidEndSepPunct{\mcitedefaultmidpunct}
{\mcitedefaultendpunct}{\mcitedefaultseppunct}\relax
\EndOfBibitem
\bibitem[Liu \latin{et~al.}(2020)Liu, Guillemeney, Choux, Ma{\^\i}tre, Ab{\'e}cassis, and Coolen]{liu2020fourier}
Liu,~J.; Guillemeney,~L.; Choux,~A.; Ma{\^\i}tre,~A.; Ab{\'e}cassis,~B.; Coolen,~L. Fourier-imaging of single self-assembled CdSe nanoplatelet chains and clusters reveals out-of-plane dipole contribution. \emph{ACS photonics} \textbf{2020}, \emph{7}, 2825--2833\relax
\mciteBstWouldAddEndPuncttrue
\mciteSetBstMidEndSepPunct{\mcitedefaultmidpunct}
{\mcitedefaultendpunct}{\mcitedefaultseppunct}\relax
\EndOfBibitem
\bibitem[Benjamin \latin{et~al.}(2020)Benjamin, Yallapragada, Amgar, Yang, Tenne, and Oron]{benjamin2020temperature}
Benjamin,~E.; Yallapragada,~V.~J.; Amgar,~D.; Yang,~G.; Tenne,~R.; Oron,~D. Temperature dependence of excitonic and biexcitonic decay rates in colloidal nanoplatelets by time-gated photon correlation. \emph{The journal of physical chemistry letters} \textbf{2020}, \emph{11}, 6513--6518\relax
\mciteBstWouldAddEndPuncttrue
\mciteSetBstMidEndSepPunct{\mcitedefaultmidpunct}
{\mcitedefaultendpunct}{\mcitedefaultseppunct}\relax
\EndOfBibitem
\bibitem[Ma \latin{et~al.}(2017)Ma, Diroll, Cho, Fedin, Schaller, Talapin, Gray, Wiederrecht, and Gosztola]{ma2017size}
Ma,~X.; Diroll,~B.~T.; Cho,~W.; Fedin,~I.; Schaller,~R.~D.; Talapin,~D.~V.; Gray,~S.~K.; Wiederrecht,~G.~P.; Gosztola,~D.~J. Size-dependent biexciton quantum yields and carrier dynamics of quasi-two-dimensional core/shell nanoplatelets. \emph{ACS nano} \textbf{2017}, \emph{11}, 9119--9127\relax
\mciteBstWouldAddEndPuncttrue
\mciteSetBstMidEndSepPunct{\mcitedefaultmidpunct}
{\mcitedefaultendpunct}{\mcitedefaultseppunct}\relax
\EndOfBibitem
\bibitem[Proppe \latin{et~al.}(2023)Proppe, Berkinsky, Zhu, {\v{S}}verko, Kaplan, Horowitz, Kim, Chung, Jun, and Bawendi]{proppe2023highly}
Proppe,~A.~H.; Berkinsky,~D.~B.; Zhu,~H.; {\v{S}}verko,~T.; Kaplan,~A.~E.; Horowitz,~J.~R.; Kim,~T.; Chung,~H.; Jun,~S.; Bawendi,~M.~G. Highly stable and pure single-photon emission with 250 ps optical coherence times in InP colloidal quantum dots. \emph{Nature Nanotechnology} \textbf{2023}, \emph{18}, 993--999\relax
\mciteBstWouldAddEndPuncttrue
\mciteSetBstMidEndSepPunct{\mcitedefaultmidpunct}
{\mcitedefaultendpunct}{\mcitedefaultseppunct}\relax
\EndOfBibitem
\bibitem[Jiang \latin{et~al.}(2021)Jiang, Roy, Claude, and Wenger]{jiang2021single}
Jiang,~Q.; Roy,~P.; Claude,~J.-B.; Wenger,~J. Single photon source from a nanoantenna-trapped single quantum dot. \emph{Nano Letters} \textbf{2021}, \emph{21}, 7030--7036\relax
\mciteBstWouldAddEndPuncttrue
\mciteSetBstMidEndSepPunct{\mcitedefaultmidpunct}
{\mcitedefaultendpunct}{\mcitedefaultseppunct}\relax
\EndOfBibitem
\bibitem[Michler \latin{et~al.}(2000)Michler, Imamo{\u{g}}lu, Mason, Carson, Strouse, and Buratto]{michler2000quantum}
Michler,~P.; Imamo{\u{g}}lu,~A.; Mason,~M.; Carson,~P.; Strouse,~G.; Buratto,~S. Quantum correlation among photons from a single quantum dot at room temperature. \emph{Nature} \textbf{2000}, \emph{406}, 968--970\relax
\mciteBstWouldAddEndPuncttrue
\mciteSetBstMidEndSepPunct{\mcitedefaultmidpunct}
{\mcitedefaultendpunct}{\mcitedefaultseppunct}\relax
\EndOfBibitem
\bibitem[Lounis \latin{et~al.}(2000)Lounis, Bechtel, Gerion, Alivisatos, and Moerner]{lounis2000photon}
Lounis,~B.; Bechtel,~H.; Gerion,~D.; Alivisatos,~P.; Moerner,~W. Photon antibunching in single CdSe/ZnS quantum dot fluorescence. \emph{Chemical Physics Letters} \textbf{2000}, \emph{329}, 399--404\relax
\mciteBstWouldAddEndPuncttrue
\mciteSetBstMidEndSepPunct{\mcitedefaultmidpunct}
{\mcitedefaultendpunct}{\mcitedefaultseppunct}\relax
\EndOfBibitem
\bibitem[Li and Lian(2017)Li, and Lian]{li2017area}
Li,~Q.; Lian,~T. Area-and thickness-dependent biexciton Auger recombination in colloidal CdSe nanoplatelets: breaking the “Universal Volume Scaling Law”. \emph{Nano letters} \textbf{2017}, \emph{17}, 3152--3158\relax
\mciteBstWouldAddEndPuncttrue
\mciteSetBstMidEndSepPunct{\mcitedefaultmidpunct}
{\mcitedefaultendpunct}{\mcitedefaultseppunct}\relax
\EndOfBibitem
\bibitem[Moerner(1997)]{moerner1997those}
Moerner,~W. Those blinking single molecules. \emph{Science} \textbf{1997}, \emph{277}, 1059--1060\relax
\mciteBstWouldAddEndPuncttrue
\mciteSetBstMidEndSepPunct{\mcitedefaultmidpunct}
{\mcitedefaultendpunct}{\mcitedefaultseppunct}\relax
\EndOfBibitem
\bibitem[Bradac \latin{et~al.}(2010)Bradac, Gaebel, Naidoo, Sellars, Twamley, Brown, Barnard, Plakhotnik, Zvyagin, and Rabeau]{bradac2010observation}
Bradac,~C.; Gaebel,~T.; Naidoo,~N.; Sellars,~M.; Twamley,~J.; Brown,~L.; Barnard,~A.; Plakhotnik,~T.; Zvyagin,~A.; Rabeau,~J. Observation and control of blinking nitrogen-vacancy centres in discrete nanodiamonds. \emph{Nature nanotechnology} \textbf{2010}, \emph{5}, 345--349\relax
\mciteBstWouldAddEndPuncttrue
\mciteSetBstMidEndSepPunct{\mcitedefaultmidpunct}
{\mcitedefaultendpunct}{\mcitedefaultseppunct}\relax
\EndOfBibitem
\bibitem[Efros and Nesbitt(2016)Efros, and Nesbitt]{efros2016origin}
Efros,~A.~L.; Nesbitt,~D.~J. Origin and control of blinking in quantum dots. \emph{Nature nanotechnology} \textbf{2016}, \emph{11}, 661--671\relax
\mciteBstWouldAddEndPuncttrue
\mciteSetBstMidEndSepPunct{\mcitedefaultmidpunct}
{\mcitedefaultendpunct}{\mcitedefaultseppunct}\relax
\EndOfBibitem
\bibitem[Frantsuzov \latin{et~al.}(2008)Frantsuzov, Kuno, Janko, and Marcus]{frantsuzov2008universal}
Frantsuzov,~P.; Kuno,~M.; Janko,~B.; Marcus,~R.~A. Universal emission intermittency in quantum dots, nanorods and nanowires. \emph{Nature Physics} \textbf{2008}, \emph{4}, 519--522\relax
\mciteBstWouldAddEndPuncttrue
\mciteSetBstMidEndSepPunct{\mcitedefaultmidpunct}
{\mcitedefaultendpunct}{\mcitedefaultseppunct}\relax
\EndOfBibitem
\bibitem[Rabouw \latin{et~al.}(2019)Rabouw, Antolinez, Brechbühler, and Norris]{rabouw2019microsecond}
Rabouw,~F.~T.; Antolinez,~F.~V.; Brechbühler,~R.; Norris,~D.~J. Microsecond blinking events in the fluorescence of colloidal quantum dots revealed by correlation analysis on preselected photons. \emph{The journal of physical chemistry letters} \textbf{2019}, \emph{10}, 3732--3738\relax
\mciteBstWouldAddEndPuncttrue
\mciteSetBstMidEndSepPunct{\mcitedefaultmidpunct}
{\mcitedefaultendpunct}{\mcitedefaultseppunct}\relax
\EndOfBibitem
\bibitem[Messin \latin{et~al.}(2001)Messin, Hermier, Giacobino, Desbiolles, and Dahan]{messin2001bunching}
Messin,~G.; Hermier,~J.-P.; Giacobino,~E.; Desbiolles,~P.; Dahan,~M. Bunching and antibunching in the fluorescence of semiconductor nanocrystals. \emph{Optics Letters} \textbf{2001}, \emph{26}, 1891--1893\relax
\mciteBstWouldAddEndPuncttrue
\mciteSetBstMidEndSepPunct{\mcitedefaultmidpunct}
{\mcitedefaultendpunct}{\mcitedefaultseppunct}\relax
\EndOfBibitem
\bibitem[D’Amato \latin{et~al.}(2023)D’Amato, Belzane, Dabard, Silly, Patriarche, Glorieux, Le~Jeannic, Lhuillier, and Bramati]{d2023highly}
D’Amato,~M.; Belzane,~L.; Dabard,~C.; Silly,~M.; Patriarche,~G.; Glorieux,~Q.; Le~Jeannic,~H.; Lhuillier,~E.; Bramati,~A. Highly photostable Zn-treated halide perovskite nanocrystals for efficient single photon generation. \emph{Nano Letters} \textbf{2023}, \emph{23}, 10228--10235\relax
\mciteBstWouldAddEndPuncttrue
\mciteSetBstMidEndSepPunct{\mcitedefaultmidpunct}
{\mcitedefaultendpunct}{\mcitedefaultseppunct}\relax
\EndOfBibitem
\bibitem[Manceau \latin{et~al.}(2018)Manceau, Vezzoli, Glorieux, Giacobino, Carbone, De~Vittorio, Hermier, and Bramati]{manceau2018cdse}
Manceau,~M.; Vezzoli,~S.; Glorieux,~Q.; Giacobino,~E.; Carbone,~L.; De~Vittorio,~M.; Hermier,~J.-P.; Bramati,~A. CdSe/CdS Dot-in-Rods Nanocrystals Fast Blinking Dynamics. \emph{ChemPhysChem} \textbf{2018}, \emph{19}, 3288--3295\relax
\mciteBstWouldAddEndPuncttrue
\mciteSetBstMidEndSepPunct{\mcitedefaultmidpunct}
{\mcitedefaultendpunct}{\mcitedefaultseppunct}\relax
\EndOfBibitem
\bibitem[Chen \latin{et~al.}(2008)Chen, Vela, Htoon, Casson, Werder, Bussian, Klimov, and Hollingsworth]{chen2008giant}
Chen,~Y.; Vela,~J.; Htoon,~H.; Casson,~J.~L.; Werder,~D.~J.; Bussian,~D.~A.; Klimov,~V.~I.; Hollingsworth,~J.~A. “Giant” multishell CdSe nanocrystal quantum dots with suppressed blinking. \emph{Journal of the American Chemical Society} \textbf{2008}, \emph{130}, 5026--5027\relax
\mciteBstWouldAddEndPuncttrue
\mciteSetBstMidEndSepPunct{\mcitedefaultmidpunct}
{\mcitedefaultendpunct}{\mcitedefaultseppunct}\relax
\EndOfBibitem
\bibitem[Mahler \latin{et~al.}(2008)Mahler, Spinicelli, Buil, Quelin, Hermier, and Dubertret]{mahler2008towards}
Mahler,~B.; Spinicelli,~P.; Buil,~S.; Quelin,~X.; Hermier,~J.-P.; Dubertret,~B. Towards non-blinking colloidal quantum dots. \emph{Nature materials} \textbf{2008}, \emph{7}, 659--664\relax
\mciteBstWouldAddEndPuncttrue
\mciteSetBstMidEndSepPunct{\mcitedefaultmidpunct}
{\mcitedefaultendpunct}{\mcitedefaultseppunct}\relax
\EndOfBibitem
\bibitem[Galland \latin{et~al.}(2012)Galland, Ghosh, Steinbr{\"u}ck, Hollingsworth, Htoon, and Klimov]{galland2012lifetime}
Galland,~C.; Ghosh,~Y.; Steinbr{\"u}ck,~A.; Hollingsworth,~J.~A.; Htoon,~H.; Klimov,~V.~I. Lifetime blinking in nonblinking nanocrystal quantum dots. \emph{Nature communications} \textbf{2012}, \emph{3}, 908\relax
\mciteBstWouldAddEndPuncttrue
\mciteSetBstMidEndSepPunct{\mcitedefaultmidpunct}
{\mcitedefaultendpunct}{\mcitedefaultseppunct}\relax
\EndOfBibitem
\bibitem[Trinh \latin{et~al.}(2020)Trinh, Minh, Ahn, Kang, and Lee]{trinh2020verification}
Trinh,~C.~T.; Minh,~D.~N.; Ahn,~K.~J.; Kang,~Y.; Lee,~K.-G. Verification of Type-A and Type-B-HC Blinking Mechanisms of Organic--Inorganic Formamidinium Lead Halide Perovskite Quantum Dots by FLID Measurements. \emph{Scientific Reports} \textbf{2020}, \emph{10}, 2172\relax
\mciteBstWouldAddEndPuncttrue
\mciteSetBstMidEndSepPunct{\mcitedefaultmidpunct}
{\mcitedefaultendpunct}{\mcitedefaultseppunct}\relax
\EndOfBibitem
\bibitem[Pierini \latin{et~al.}(2020)Pierini, d’Amato, Goyal, Glorieux, Giacobino, Lhuillier, Couteau, and Bramati]{pierini2020highly}
Pierini,~S.; d’Amato,~M.; Goyal,~M.; Glorieux,~Q.; Giacobino,~E.; Lhuillier,~E.; Couteau,~C.; Bramati,~A. Highly photostable perovskite nanocubes: toward integrated single photon sources based on tapered nanofibers. \emph{ACS photonics} \textbf{2020}, \emph{7}, 2265--2272\relax
\mciteBstWouldAddEndPuncttrue
\mciteSetBstMidEndSepPunct{\mcitedefaultmidpunct}
{\mcitedefaultendpunct}{\mcitedefaultseppunct}\relax
\EndOfBibitem
\bibitem[Kim \latin{et~al.}(2019)Kim, Jung, Ham, Chung, and Kim]{kim2019elucidation}
Kim,~T.; Jung,~S.~I.; Ham,~S.; Chung,~H.; Kim,~D. Elucidation of photoluminescence blinking mechanism and multiexciton dynamics in hybrid organic--inorganic perovskite quantum dots. \emph{Small} \textbf{2019}, \emph{15}, 1900355\relax
\mciteBstWouldAddEndPuncttrue
\mciteSetBstMidEndSepPunct{\mcitedefaultmidpunct}
{\mcitedefaultendpunct}{\mcitedefaultseppunct}\relax
\EndOfBibitem
\bibitem[Yuan \latin{et~al.}(2018)Yuan, G{\'o}mez, Kirkwood, Boldt, and Mulvaney]{yuan2018two}
Yuan,~G.; G{\'o}mez,~D.~E.; Kirkwood,~N.; Boldt,~K.; Mulvaney,~P. Two mechanisms determine quantum dot blinking. \emph{ACS nano} \textbf{2018}, \emph{12}, 3397--3405\relax
\mciteBstWouldAddEndPuncttrue
\mciteSetBstMidEndSepPunct{\mcitedefaultmidpunct}
{\mcitedefaultendpunct}{\mcitedefaultseppunct}\relax
\EndOfBibitem
\bibitem[Li \latin{et~al.}(2024)Li, Biesterfeld, Klepzig, Yang, Ngo, Addad, Rakow, Guan, Rugeramigabo, Biadala, \latin{et~al.} others]{li2024polarized}
Li,~P.; Biesterfeld,~L.; Klepzig,~L.; Yang,~J.; Ngo,~H.~T.; Addad,~A.; Rakow,~T.~N.; Guan,~R.; Rugeramigabo,~E.~P.; Biadala,~L.; others Polarized sub-meV Photoluminescence in 2D PbS Nanoplatelets at Cryogenic Temperatures. \emph{arXiv preprint arXiv:2405.19821} \textbf{2024}, \relax
\mciteBstWouldAddEndPunctfalse
\mciteSetBstMidEndSepPunct{\mcitedefaultmidpunct}
{}{\mcitedefaultseppunct}\relax
\EndOfBibitem
\end{mcitethebibliography}

\end{document}



\section{S1. Synthesis of the CdSe/ZnS NPLs}

\subsection{S1.1 Chemicals}

Octadecene (ODE) (Sigma-Aldrich, $90\%$), cadmium acetate dihydrate (\ch{Cd(Ac)_2}$\cdot$ 2\ch{H_2}O)
(Sigma-Aldrich, $98\%$), cadmium oxide (CdO) (Strem $99.99\%$),
myristic acid (Sigma-Aldrich, $99\%$), selenium powder (Se) (Strem Chemicals $99.99\%$), oleic acid (OA) (Sigma- Aldrich, $90\%$), 
oleylamine (OLA, Acros, $80-90\%$), trioctylphosphine (TOP) (Alfa Aesar, $90\%$), octanethiol (Sigma Aldrich, $98.5\%$), 
n-hexane (VWR, $99\%$), methanol (MeOH) (VWR, $99.8\%$) and ethanol absolute (VWR, $99.8\%$) are used. All chemicals are used as received. 

\subsection{S1.2 1 M TOP-Se.} In the glovebox, \SI{1.58}{\gram} of Se powder was added to \SI{20}{\milli \liter} of TOP, and the solution was stirred overnight at room temperature until the complete dissolution of the Se powder. The solution was then stored in a glovebox.

\subsection{S1.3 Synthesis of Cadmium Myristate.} \SI{2.56}{\gram} (\SI{20}{\milli \mole}) of CdO and \SI{11}{\gram} (\SI{50}{\milli \mole}) of myristic acid were loaded into a \SI{50}{\milli \liter} three-neck flask. The mixture was heated to \SI{80}{\celsius} and degassed for \SI{30}{\minute}. Under an argon flow, the solution was kept heating at \SI{200}{\celsius} until becoming colorless. During the cooling step, \SI{30}{\milli \liter} of MeOH was added at \SI{60}{\celsius} in order to solubilize the excess myristic acid. The obtained white solid was precipitated 5 times with MeOH using a centrifuge tube. The final product was dried at \SI{70}{\celsius} under vacuum overnight.

\subsection{S1.4 Synthesis of Cadmium Acetate.} The as-purchased cadmium acetate dihydrate is crushed and then dried under vacuum at \SI{70}{\celsius} overnight.

\subsection{S1.5 Nanoplateletes Synthesis} 

 \textbf{Rectangular 8 nm x 20 nm 4 ML CdSe core NPLs.} The 4 ML zinc-blende CdSe NPLs were synthesized following Ithurria et al. procedure\cite{ithurria2012colloidal}. In a \SI{25}{\milli \liter} three-neck flask, \SI{170}{\milli \gram} of \ch{Cd(myristate)_2}, \SI{12}{\milli \gram} of Se powder and \SI{15}{\milli \liter} of ODE were loaded and degassed under vacuum for \SI{30}{\minute} at room temperature. The mixture was then heated to \SI{240}{\celsius} under an argon flow. At around \SI{200}{\celsius}, \SI{90}{\milli \gram} of dried \ch{Cd(Ac)2} was quickly added to the intense-colored orange solution. The reaction was stopped when the \SI{512}{\nano \meter} absorption was stabilized (around \SI{20}{\minute} after the addition of \ch{Cd(Ac)2}). During the cooling step, \SI{150}{\micro \liter} of OA was injected into the solution at \SI{150}{\celsius}. After cooling, the NPLs were purified twice with \SI{15}{\milli \liter} of hexane and \SI{15}{\milli \liter} of ethanol by centrifuging at 6000 rpm for \SI{5}{\minute}.
\newline
\textbf{ CdSe/ZnS core-shell NPLs.} 
The synthesis is modified from a method by Kelestemur et al\cite{kelestemur2019colloidal}. The amount of CdSe NPLs used for the synthesis was equivalent to \SI{10}{\milli \liter} of CdSe NPLs solution with an optical density of 0.2 at the highest excitonic peak, considering a \SI{1}{\centi \meter} optical path length. The CdSe NPLs were precipitated with ethanol and then dissolved in \SI{10}{\milli \liter} of ODE in a \SI{25}{\milli \liter} three-neck flask. Before the mixture was degassed at room temperature for \SI{30}{\minute}, \SI{88}{\milli \gram} of zinc acetate (\SI{0.48}{\milli \mole}) and \SI{1}{\milli \liter} of OA were loaded. The second degas process was conducted at \SI{85}{\celsius} for \SI{30}{\minute}. Under argon, \SI{1}{\milli \liter} of oleylamine was added to the mixture and the flask was heated up to \SI{300}{\celsius}. At \SI{160}{\celsius}, we began to inject a thiol solution (\SI{88}{\micro \liter} of octanethiol in \SI{5}{\milli \liter} of ODE) at a rate of 10 mL/h, when the solution became dark red, the injection rate was decreased to 4 mL/h. The reaction was stopped when the PL feature reached the expected wavelength value. The coating is typically \SI{1}{\nano \meter} thick. The synthesized core-shell NPLs were diluted in \SI{10}{\milli \liter} of hexane in a \SI{50}{\milli \liter} centrifuge tube and were precipitated with the addition of \SI{10}{\milli \liter} ethanol at 6000 rpm for \SI{5}{\minute}. After repeating once more the purification step, the final NPLs were suspended in \SI{3}{\milli \liter} hexane. \newline

\section{S2. Wide-field and confocal microscopy at single-particle level}

\begin{figure*}[ht!]
     \includegraphics[width=140 mm]{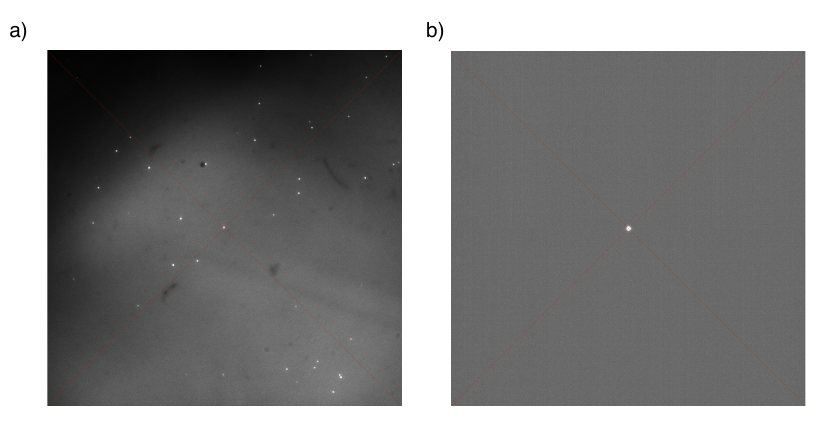}
     \caption{ Optical characterization at single-particle level: a) Wide-field microscopy image of CdSe/ZnS NPLs. b) Confocal microscopy image of an individual CdSe/ZnS NPL.}
   \label{fig:FIGS1}
\end{figure*}

The synthesized CdSe/ZnS NPLs show good dispersion stability without noticeable aggregation, as shown in Figure \ref{fig:FIGS1}a. This enabled confocal spectroscopy at single particle level to probe their single-photon emission, as visible in Figure \ref{fig:FIGS1}b.

\section{S3. PL time-trace}

Figure S1 shows the photoluminesce time trace over the all 600 s of acquisition time, relative to Figure 3a. The detected photons are binned with a 5 ms bin. 

\begin{figure*}[ht!]
     \includegraphics[width=140 mm]{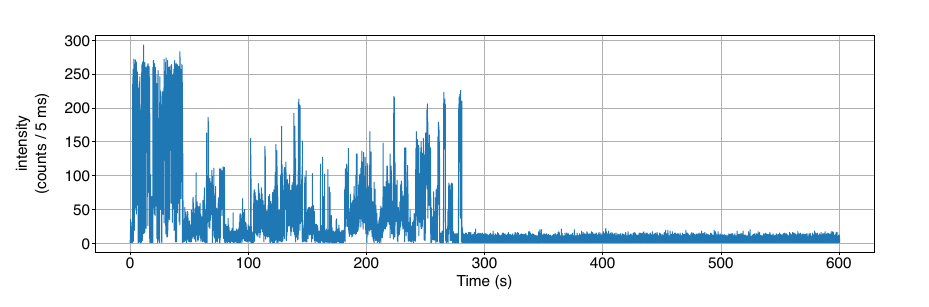}
     \caption{ PL time trace of a single NPL over 600 s  with a bin time of 5ms.}
   \label{fig:FIGS2}
\end{figure*}

\bibliography{Manuscript}